\documentclass[fleqn,10pt]{wlscirep}
\usepackage[utf8]{inputenc}
\usepackage[T1]{fontenc}
\usepackage{threeparttable}
\title{Location-Based Service (LBS) Data Quality Metrics and Effects on Mobility Inference}

\author[1]{Xinhua Wu}
\author[1]{Yanchao Wang}
\author[2]{Ekin Ugurel}
\author[2]{Cynthia Chen}
\author[3]{Shuai Huang}
\author[1,*]{Qi R. Wang}

\affil[1]{Northeastern University, Department of Civil and Environmental Engineering, Boston, MA 02115, United States}
\affil[2]{University of Washington, Department of Civil and Environmental Engineering, Seattle, WA 98195, United States}
\affil[3]{University of Washington, Department of Industrial and Systems Engineering, Seattle, WA 98195, United States}

\affil[*]{Corresponding author: q.wang@northeastern.edu}

\begin{abstract}
Today, GPS-equipped mobile devices are ubiquitous, and they generate Location-Based Service (LBS) data, which has become a critical resource for understanding human mobility. However, inherent limitations in LBS datasets, primarily characterized by discontinuity and sparsity, may introduce significant biases in representing individual movement patterns. This study develops data quality metrics for LBS data, examines their disparities among different populations, and quantifies their effects on inferred individual movement, stays in particular, in the Boston Metropolitan Area. 
We find that data from higher-income, more educated, and predominantly white census block groups (CBGs) show higher sampling rates but paradoxically lower data quality. This contradiction may stem from greater privacy awareness in these communities. Additionally, we propose a new framework to resample LBS data and quantitatively evaluate the inferential biases associated with data of varying quality. This versatile framework can analyze the impacts originating from different data processing workflows with LBS data. Using linear regression models with clustered standard error, we assess the impact of data quality metrics on inferring the number of stay points. The results show that better data quality, characterized by the number of observations and temporal occupancy, can significantly reduce the bias when calculating the stay points of an individual. The introduction of additional data quality metrics into the regression model can further explain the bias. Overall, this study provides insights into how data quality can influence our understanding of human mobility patterns, highlighting the importance of carefully handling LBS data in research.
\end{abstract}
\begin{document}

\flushbottom
\maketitle
\thispagestyle{empty}

% keywords: LBS Data, Data Quality Metrics, Disparities in Data Quality, Data Biases

% \noindent Please note: Abbreviations should be introduced at the first mention in the main text – no abbreviations lists. Suggested structure of main text (not enforced) is provided below.

\section*{Introduction}

With the widespread adoption of GPS-equipped mobile devices, an immense volume of location-based service data has been generated. This type of data set allows for mapping out the mobility patterns of a significant number of individuals within a given region, promising to reduce or eliminate the need for traditional, labor-intensive survey efforts. Large-scale LBS data can also potentially allow researchers and practitioners to detect shifts in mobility patterns over time, determine the factors that influence such changes, and help design appropriate policy changes or interventions. Numerous studies have used LBS data for a variety of applications, including understanding individual mobility patterns \cite{wang2018urban, gonzalez_understanding_2008, he2023forecasting}, estimating origin-destination traffic flow \cite{calabrese_real-time_2011, paipuri_estimating_2020}, predicting pandemic
spreading \cite{frias-martinez_estimation_2012, wang_identifying_2022} and facilitating urban planning strategies \cite{sulis_using_2018, wang_defining_2023}.

However, because LBS data originates from users’ interactions with their smart devices, it is often discontinuous and sparse across time and space \cite{chen_promises_2016, he2022percolation}. Thus, the individual mobility patterns inferred from such data, such as stayed locations, travel distances, duration for their visits, origins, and destinations, is likely incomplete \cite{kilic_missingness_2017, barnett_inferring_2020}.
The incompleteness could be due to several reasons. Firstly, users have control over whether or not to enable their location services. They can disable data collection entirely or restrict individual applications from accessing their location for reasons ranging from privacy concerns to battery conservation. Studies have found a correlation between the willingness to allow location tracking by apps and perceived user-end benefits \cite{kim_willingness_2019}. Users are less likely to consent to provide continuous location data for an application unless they recognize distinct advantages. Secondly, the built environment also contributes to the discontinuity of LBS data. For instance, the urban canyon effect, caused by high-rise buildings and tunnels interfering with GPS signals in urban environments, can result in observations moving at high speed within a short interval (oscillations) \cite{groves_shadow_2011, Chen201858}. Lastly, the operational constraints of LBS data collection play a role. Some operating systems restrict background app activities, which can limit how often location data is transmitted \cite{zhou_demystifying_2020}. Even if a device continuously generates location data, it might only be sent to data collectors at certain intervals, resulting in a less dense dataset.

The discontinuity and sparsity of LBS data can lead to inaccurate and biased representations of individual movements, including but not limited to the extent of personal travel \cite{chen_promises_2016} (i.e., the number of trips calculated and the associated spatial extent), the inferred duration of activities \cite{mccool_maximum_2022}, and the portrayal of users’ interactions with the built environment.
% \cite{chen_promises_2016} found that missing records in LBS data can substantially distort the estimation of visitation patterns. \cite{mccool_maximum_2022} revealed that LBS data with long record gaps lead to a downward bias on travel distance, movement events, and radius of gyration. 
Additionally, the biases can be exacerbated in certain demographics and populations. Studies show that LBS data may not accurately represent minority groups or may skew towards more affluent users, leading to disparities in data quality among different populations \cite{wesolowski2012heterogeneous, mohorko2013coverage}. 
The biases inherent in the quality of LBS data can lead to the misunderstanding of mobility, hindering the formulation of unbiased conclusions and the development of fair and effective policies. 
However, existing studies often overlook the differences in data quality among individuals and seldom discuss the potential biases caused by underlying data quality. It is imperative to rigorously analyze the impact of LBS data quality on individual human mobility.

To fill this gap, this research investigates the quality of LBS data and its consequent effects on individual mobility metrics. We examined the LBS data quality in the Boston Metropolitan Area, uncovering a notable correlation between data quality and demographic factors. Moreover, we introduced a resampling-based LBS bias assessment framework to quantitatively evaluate the inferential bias in mobility that might arise from LBS data. This framework isolates high-quality LBS data, employs resampling to create incomplete datasets with varied quality, and analyzes their biases from established ground truth. To mitigate the autocorrelation introduced by resampling, we applied clustered standard errors which remedy the violation of the independence assumption in regression. We assessed the influence of different data quality metrics on the inference of the number of stay points. 

We found disparities in data quality from different population groups. Census block groups (CBGs) characterized by higher incomes, better educational outcomes, and a greater white portion of the population exhibit higher LBS sampling rates (the ratio of users to the overall population), but paradoxically, the data quality appears lower. This discrepancy might be linked to the greater prevalence of smartphone use in these areas and an increased awareness of preserving data privacy. 
Additionally, the regression results showed that an increase of 100 in the number of observations per day for a user would help reduce the bias by 0.2 stay points. Increasing temporal occupancy by 10 would help reduce the bias by 1.2. Improvements in other metrics, such as the Maximum record gap, also contribute to reducing errors in inferring mobility, which provides a general reference for the potential bias of LBS data for understanding human mobility.

% The rest of the paper is organized in the following sections: Section \ref{sec:metric} introduces the datasets and metrics used in the study. Section \ref{sec:quality} discusses the quality of the LBS data and its inequity across different demographic groups. Section \ref{sec:method} describes the resampling-based LBS bias assessment framework for estimating the inferential bias. Section \ref{sec:result} applies the proposed framework to real-world LBS data and provides insights into the relationship between data quality and bias. The paper concludes with Section \ref{sec:conclusion}, summarizing the findings and implications of the study.

\section*{Dataset and metrics}

\subsection*{Mobility and Demographic Datasets}

In this study, we utilized LBS data from the Boston Metropolitan Area provided by Spectus Inc. This dataset spans from January 1 to February 28, 2020. Each record in the dataset includes a user's geographical coordinates, specifically latitude and longitude, along with a precise timestamp when the data was collected. Additionally, the dataset includes information on spatial precision, characterized as the radius around the exact GPS location. Within this radius, there is a 95\% probability that it encompasses the actual location of the user.
In addition to the LBS dataset, we also leverage the 5-year American Community Survey (ACS) data spanning from 2015 to 2019 as our demographic reference, aiming to understand how LBS data quality varies across different demographic groups.

\subsection*{Data Quality Metrics}

Table \ref{tab:metrics} shows the data quality metrics employed in this study and their definitions. These metrics capture different dimensions of data characteristics, including frequency and regularity of observations, temporal intervals, spatial accuracy, and the temporal distribution of the data points, providing a multifaceted view of the quality and coverage of the data collected from each device. In this study, we gauge the quality of LBS data by analyzing a user's 24-hour data record. By examining this data through the lens of our defined metrics, we obtain a comprehensive understanding of the data's quality across multiple dimensions.

\begin{table*}[ht]
    \caption{Description of Data Quality Metrics}
    \label{tab:metrics}
    \centering
    \small
    \renewcommand{\arraystretch}{1.1}
    \begin{threeparttable}
    \begin{tabular}{l@{\hspace{1.5cm}}p{9cm}}
        \toprule
        \textbf{Metrics} & \textbf{Definition} \\ \midrule
        Number of observations   & The number of records from a device \\ \midrule
        Temporal occupancy \cite{sulis_using_2018} & The number of 30-min slots in which a device is observed at least once, ranging from 0 to 48\\ \midrule
        Maximum record gap & The largest duration between two consecutive records (in minute) \\ \midrule
        Percentage of high accuracy observations  & Percentage of observations with spatial accuracy lower than 100 meters (\%) \\ \midrule
        Temporal burstiness \cite{goh_burstiness_2006} & Distributional pattern for observational interval in minutes [-1,1]; If evenly distributed, -1; if randomly distributed, 0; if distributed as clusters, 1 \\
        \bottomrule
    \end{tabular}
    \end{threeparttable}
\end{table*}

\section*{LBS Data Quality}
In this section, we explore the quality of the existing LBS data based on these metrics.

\subsection*{Qualification criterion and CBG segmentation}

We developed three criteria to assess data quality, structured in an inclusive relationship. \textbf{Criterion 1} stands out as the most rigorous, demanding a high degree of data completeness and density, whereas \textbf{Criterion 3} is the most lenient. Note that the data conforming to \textbf{Criterion 1} satisfy \textbf{Criterion 2}, and similarly, the data adhering to \textbf{Criterion 2} fulfill \textbf{Criterion 3}. These criteria are:

\begin{itemize}
\item \textbf{Criterion 1} : \textit{temporal occupancy} $\geq$ 40 \& \textit{maximum record gap} $\leq$ 40 min \& \textit{number of records} $\geq$ 300
\item \textbf{Criterion 2} : \textit{temporal occupancy} $\geq$ 20 \& \textit{maximum record gap} $\leq$ 120 min \& \textit{number of records} $\geq$ 100
\item \textbf{Criterion 3} : \textit{temporal occupancy} $\geq$ 10 \& \textit{maximum record gap} $\leq$ 480 min \& \textit{number of records} $\geq$ 20
\end{itemize}

These three criteria set different benchmarks for data quality related to continuity and sparsity. The percentage of days meeting these criteria indicates LBS data quality within a specific region.

Furthermore, we investigated the relationship between the quality of LBS data and demographic factors. Specifically, we estimated the census block group (CBG) of each user's home location based on their most frequent location during the nighttime (10:00 PM - 7:00 AM), as indicated by the LBS data \cite{phithakkitnukoon_socio-geography_2012,csaji_exploring_2013}. We then segmented these CBGs into distinct groups based on their demographic characteristics, including income, education level, and race. The detailed strategy for the segmentation is outlined as follows:

\textbf{1) Income-based segmentation.} We classified the CBGs into five quintiles (A1-A5) based on the median household income of each CBG.

\textbf{2) Education-based segmentation.} Similarly, the CBGs were segmented into five quintiles (B1-B5) according to the percentage of residents with a bachelor's degree or higher. 

\textbf{3) Race-based segmentation.} For analyzing the demographic influence of race on LBS data quality, we defined five categories of CBGs: C1 (Majority White), C2 (Majority Black), C3 (Majority Asian), C4 (Majority Hispanic), and C5 (Majority Mixed). A CBG was classified into one of the first four categories (C1-C4) if a single race constituted over 50\% of its population. Conversely, CBGs with no single racial group exceeding 50\% were categorized as C5 (Mixed), reflecting a more diverse demographic composition.

\subsection*{Inequity in the quality of LBS data}
We compared LBS data quality across these segmentation groups and presented the results in Table \ref{tab:cbgs}. Overall, the rates of high-quality data are low across all groups. When considering the cumulative data from all CBGs, among all 25,028,921 days, only 283,083 days satisfy \textbf{Criterion 1}. The proportion of qualified days does not exceed 1.15\%, which indicates that a considerable portion of user movement remains unrecorded in current LBS data.
The reliance on such data without due caution may introduce significant inferential biases in representing individual mobility patterns.

Moreover, there is a significant disparity in LBS data across different demographic areas. In terms of the sampling rates, areas with higher median incomes seem to have a higher proportion of the population contributing to LBS data, a trend also evident in areas with higher education levels and predominantly white communities. For example, in CBGs with a higher rate of bachelor's degree holders (B5), the sampling rate reaches 11.11\%, compared to only 6.96\% in CBGs with lower bachelor's degree rates (B1). Similarly, the 2,693 predominantly white CBGs (C1) have a sampling rate of 9.30\%, while the 139 predominantly black CBGs (C2) have a much lower rate of 5.79\%. This may be related to the higher number of smart devices per capita in these areas and the frequency of use of specific applications.

However, a higher sampling rate does not suggest a superior data quality. In higher-income areas, communities with higher levels of education, and predominantly white neighborhoods, the proportion of high-quality LBS data (\textbf{Criterion 1}) is notably lower. The qualified rate exhibits a marked decrease with the ascending income and educational levels of the CBGs. CBGs with predominantly black populations also demonstrate the highest rate of qualified data among all race-segmented CBG groups. This discrepancy may be attributed to a greater emphasis on privacy and more effective management of application permissions in certain communities.
% The statistical differences in the sampling rate and quality of LBS data across various demographic areas were assessed using the Mann–Whitney U test \cite{mann1947test}, and key results are shown in Table \ref{tab:test} in the appendix.

\begin{table*}[!t]
\caption{Comparative Analysis of LBS Data Quality Across CBGs with Varied Demographic Profiles (Boston metropolitan area, January 1 to February 28, 2020)}
\label{tab:cbgs}
\centering
\small
\begin{threeparttable}
\begin{tabular}{c@{\hspace{15pt}}ccccc@{\hspace{15pt}}c}
\toprule
\multicolumn{7}{c}{\textbf{Income-segmented CBGs}} \\
\midrule
  & A1  & A2 & A3 & A4 & A5 \tnote{$\dagger$} & Total \tnote{c} \\
\midrule
Number of CBGs & \text{656} & \text{655} & \text{656} & \text{654} & \text{655} & \text{3,276} \\
Total population & \text{909,413} & \text{935,676} & \text{950,352} & \text{949,155} & \text{956,609} & \text{4,701,205} \\
Median household Income (average) & \text{44,043} & \text{72,657} & \text{95,560} & \text{118,187} & \text{169,435} & \text{99,953} \\
LBS sampling rate (\%) \tnote{a} & \phantom{**}\text{7.79}\phantom{**} & \phantom{**}\text{7.73}*\phantom{*} & \phantom{**}\text{8.76}** & \phantom{**}\text{10.51}** & \text{8.79} & \text{8.73} \\
Qualified rate (\%) (\textbf{Criterion 1}) \tnote{b} & \phantom{**}\text{1.19}** & \phantom{**}\text{1.40}** & \phantom{**}\text{1.24}** & \phantom{**}\text{1.02}** & \text{0.94} & \text{1.15} \\
Qualified rate (\%) (\textbf{Criterion 2}) \tnote{b} & \phantom{**}\text{15.50}** & \phantom{**}\text{17.74}** & \phantom{**}\text{17.38}** & \phantom{**}\text{15.64}** & \text{18.21} & \text{16.87} \\
Qualified rate (\%) (\textbf{Criterion 3}) \tnote{b} & \phantom{**}\text{31.11}** & \phantom{**}\text{33.43}** & \phantom{**}\text{31.62}** & \phantom{**}\text{27.80}\phantom{**} & \text{31.37} & \text{30.87} \\
\toprule
\multicolumn{7}{c}{\textbf{Education-segmented CBGs}} \\
\midrule
  & B1 & B2 & B3 & B4 & B5 \tnote{$\dagger$} & Total \tnote{c} \\
\midrule
Number of CBGs & \text{681} & \text{681} & \text{681} & \text{681} & \text{681} & \text{3,405} \\
Total population & \text{988,549} & \text{1,016,947} & \text{1,015,905} & \text{953,877} & \text{859,065} & \text{4,834,343} \\
Bachelor's or higher degree rate (\%) & \text{10.38} & \text{23.15} & \text{33.69} & \text{44.86} & \text{60.56} & \text{33.61} \\
LBS sampling rate (\%) \tnote{a} & \phantom{**}\text{6.96}*\phantom{*} & \phantom{**}\text{8.81}** & \phantom{**}\text{8.95}** & \phantom{**}\text{8.32}** & \text{11.11} & \text{8.78} \\
Qualified rate (\%) (\textbf{Criterion 1}) \tnote{b} & \phantom{**}\text{1.46}** & \phantom{**}\text{1.54}** & \phantom{**}\text{1.26}** & \phantom{**}\text{1.02}** & \text{0.48} & \text{1.13} \\
Qualified rate (\%) (\textbf{Criterion 2}) \tnote{b} & \phantom{**}\text{18.11}** & \phantom{**}\text{19.50}** & \phantom{**}\text{18.67}** & \phantom{**}\text{17.94}** & \text{10.09} & \text{16.69} \\
Qualified rate (\%) (\textbf{Criterion 3}) \tnote{b} & \phantom{**}\text{35.47}** & \phantom{**}\text{35.12}** & \phantom{**}\text{33.26}** &\phantom{**}\text{31.97}**& \text{19.44} & \text{30.66} \\
\toprule
\multicolumn{7}{c}{\textbf{Race-segmented CBGs}} \\
\midrule
  & C1 \tnote{$\dagger$} (White)  & C2 (Black) & C3 (Asian) & C4 (Hispanic) & C5 (Mixed) & Total \tnote{c} \\
\midrule
Number of CBGs & \text{2,693} & \text{139} & \text{29} & \text{170} & \text{373} & \text{3,404} \\
Total population & \text{3,846,221} & \text{177,393} & \text{34,235} & \text{254,600} & \text{520,914} & \text{4,833,363} \\
LBS sampling rate (\%) \tnote{a} & \text{9.30} & \phantom{**}\text{5.79}** & \phantom{**}\text{7.06}\phantom{**} & \phantom{**}\text{6.54}** & \phantom{**}\text{7.13}** & \text{8.78} \\
Qualified rate (\%) (\textbf{Criterion 1}) \tnote{b} & \text{1.12} & \phantom{**}\text{1.55}** & \phantom{**}\text{0.89}\phantom{**} & \phantom{**}\text{1.12}\phantom{**} & \phantom{**}\text{1.19}\phantom{**} & \text{1.13} \\
Qualified rate (\%) (\textbf{Criterion 2}) \tnote{b} & \text{16.91} & \phantom{**}\text{18.49}*\phantom{*} & \phantom{**}\text{14.16}*\phantom{*} & \phantom{**}\text{13.89}** & \phantom{**}\text{15.50}** & \text{16.69}\\
Qualified rate (\%) (\textbf{Criterion 3}) \tnote{b} & \text{30.54} & \phantom{**}\text{36.76}\phantom{**} & \phantom{**}\text{30.97}\phantom{**} & \phantom{**}\text{28.80}*\phantom{*} & \phantom{**}\text{30.93}** & \text{30.66} \\
\bottomrule
\end{tabular}%
\begin{tablenotes}
\item[a] The ratio of the number of users included in the LBS data to the total population of the group.
\item[b] The ratio of the number of days to the total days in the LBS data according to the specific criterion.
\item[c] Due to the missing demographic data, the study CBGs and the LBS data vary slightly across the three segmentation analyses.
\item[$\dagger$] The Mann–Whitney U test \cite{mann1947test} assesses the statistical significance of differences compared to the base groups. * $p < 0.05$, ** $p < 0.01$
\end{tablenotes}
\end{threeparttable}
\end{table*}

\section*{Resampling-based LBS bias assessment}
\subsection*{Framework overview}
\begin{figure*}[ht]
\centering
\includegraphics[width=6.8in]{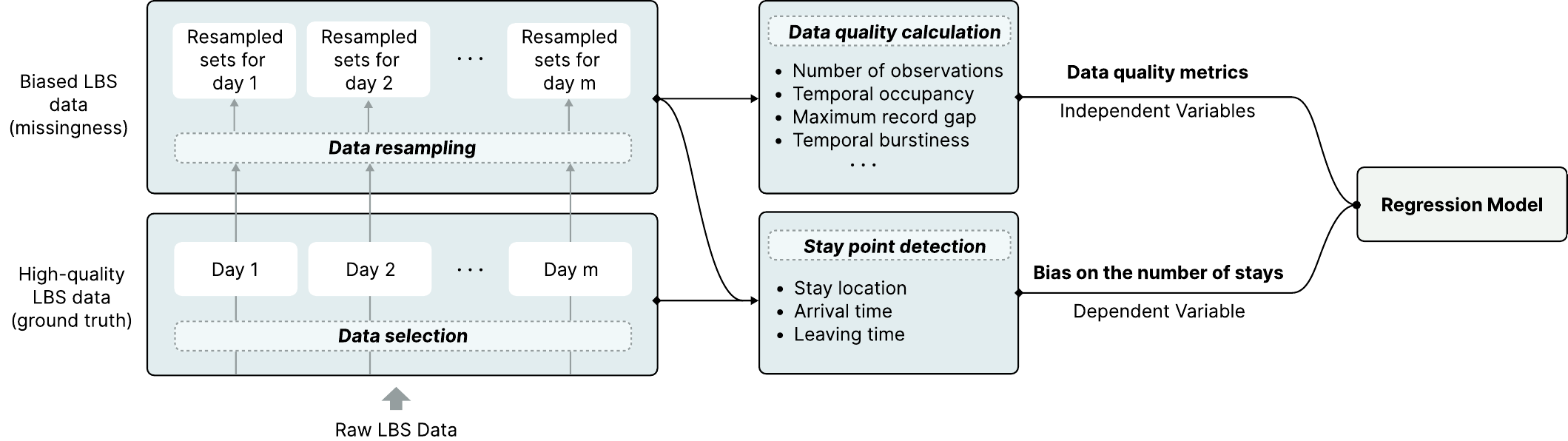}
\caption{Resampling-based framework to assess the potential bias in LBS data}
\label{fig:framework}
\end{figure*}

In this section, we explore potential effects on individual mobility information inferred from using LBS data of varying qualities. A significant challenge is the lack of ground truth in most LBS datasets, as the actual activities of users is often unknown. To tackle this issue, we proposed a resampling-based framework to assess the bias in LBS data, illustrated in Figure \ref{fig:framework}. This framework primarily focuses on isolating high-quality LBS data and using its inferred outcomes as a proxy for ground truth. We then employ resampling, a method of generating incomplete datasets by varying the quality parameters of LBS data, to create resampled datasets. The next step involves analyzing the deviation of inferences from these resampled datasets of varying levels of quality in comparison to our established ground truth. Finally, we utilize a regression model to quantitatively analyze the relationship between LBS data quality and the corresponding bias in mobility inference. It's important to note that the framework is highly flexible, where each module, including data selection, data resampling, data quality calculation, stay point detection, and regression model, can be adjusted according to the characteristics of the data and the LBS data processing workflow.

\subsection*{Data Selection and Resampling}
The resampling framework starts with data selection (Figure \ref{fig:framework} left bottom). The purpose of data selection is to identify high-quality data based on certain data quality metrics, assuming that using these high-quality datasets for inferring human mobility is nearly unbiased. 
Given the rarity of continuous, long-term, high-quality data in practical LBS datasets, the 24-hour records from a user are employed as the unit for data selection. This selection process balances data quality and the accessible volume of data, as excessively rigorous selection criteria might result in a scarcity of viable samples.

After selecting the high-quality data from an LBS data set, we can start data resampling (Figure \ref{fig:framework} left top). Data resampling involves resampling these carefully selected days through various methods, such as 10 or 20\% random downsampling, to simulate LBS data of different qualities as found in real-world scenarios. The resampling rates can be driven by the data quality observed in empirical LBS data. Diverse resampling methods can be implemented within the same day to augment the variability and richness of the dataset. 

\subsection*{Data quality calculation and stay point detection}
\noindent For each resampled day, we then calculate data quality metrics (Figure \ref{fig:framework} middle top). Beyond the metrics specified in Table \ref{tab:metrics}, additional relevant data quality metrics can be added. These metrics will act as inputs for a particular regression model. When employing a linear model, it's essential to consider and address the multicollinearity among these data quality metrics.

Using the high-quality and resampled data sets, we also obtain metrics related to mobility patterns (Figure \ref{fig:framework} middle bottom). Here, we use stay points as the metric for assessing bias, since they often serve as the starting point and foundation of mobility studies \cite{barbosa2018human,wang2023percolation}. Stay point detection needs to be conducted on each high-quality day and resampled day.
The algorithms and hyperparameters for stay point detection \cite{sahr_geodesic_2003, pappalardo_scikit-mobility_2019, guan_mobility_2022} can be tailored according to the processing workflow of LBS data.

\subsection*{Regression model}

We develop an ordinary least squares (OLS) regression model to investigate the relationship between data quality and LBS bias. The independent variables $X$ are the data quality metrics calculated with resampled LBS data. Our dependent variable, the bias $Y$, is defined as the discrepancy between the number of stay points estimated with resampled data and the ground truth determined by the original high-quality data. 

Before running the OLS model, it is important to note that creating multiple resampled data sets by resampling on the same days can introduce autocorrelation into the error term of our regression model \cite{liang_longitudinal_1986}. This potentially violates the independence assumption inherent in OLS regression. The correlation of error terms across different samples could lead to inaccurate estimation of the standard errors of our estimates, potentially leading to spurious inferences. To mitigate this issue, we introduce clustered standard errors, which remedy the violation of the independence assumption \cite{colin_cameron_practitioners_2015}. The concept behind clustered standard errors lies in the assumption that observations within the same cluster may not be independent, but clusters themselves are independent of each other. In this study, multiple resampled days from the same day form a cluster. 

Specially, we start with a standard OLS model:

\begin{equation}
\label{eq:ols1}
Y=X \beta+e
\end{equation}

\noindent where $Y$ is a vector of inferential bias on the number of stay points, $X$ is an $n * m$ matrix of data quality metrics, $\beta$ is an $m * 1$ vector of unknown parameters, and $e$ is an $m * 1$ vector of residuals.

Since residuals exhibit correlation within clusters, we turn to the heteroscedasticity-robust standard error, which provides a consistent estimate of $V(\hat{\beta})$ under heteroscedasticity:

\begin{equation}
\label{eq:ols2}
V(\hat{\beta})=\left(X^{\prime} X\right)^{-1} X^{\prime} \Omega X\left(X^{\prime} X\right)^{-1}
\end{equation}

Here, $\Omega$ is the covariance matrix of residuals. The covariance values within each cluster are unrestricted, but the covariance is assumed to be zero between clusters.

\section*{Case study}

\subsection*{Experiment setting}
In our experiments, we carefully selected high-quality data of 24-hour records from 132 users, adhering to the following criterion: \textit{temporal occupancy} $=$ 48, \textit{maximum record gap} $\leq$ 20 min, and \textit{number of records} $\geq$ 500. This helped us obtain unbiased mobility behaviors as the ground truth. Each user was limited to contributing only one day of data to avoid the potential bias that might arise from including multiple days from the same individual. Having identified these high-quality data, we proceeded to randomly resample them at various rates, ranging from 1\% to 90\%. To minimize the inherent randomness associated with resampling, this process was repeated 10 times each day, reducing the likelihood of anomalies that could result from a single, non-representative resample.

For our analysis, we employed the five data quality metrics listed in Table \ref{tab:metrics}. Also, we utilized the built-in function provided by the Scikit-mobility library to detect the stay points \cite{pappalardo_scikit-mobility_2019}. With other parameters remaining default, we adjusted the “no\_data\_for\_minutes” parameter to 30, indicating that any interval exceeding 30 minutes without data would not be recognized as a stay point.

\subsection*{Results}

The results from the regression models, as shown in Table \ref{tab:res}, incorporate different combinations of independent variables to assess their influence on the bias in the number of stay points detected per day. For example, a bias of $-1$ indicates that one stay point within that day were not correctly identified. The presence of negative constant terms across all models suggests that, in general, there is a tendency to underestimate the number of stay points due to the incompleteness of LBS data. The bias often appears as a negative number because resampled incomplete data typically leads to the algorithm's failure to recognize some stay points accurately. 

Here, \textbf{Model 1} leverages only two principal data quality metrics, \textit{number of observations} and \textit{temporal occupancy}, to provide a substantive explanation for the observed bias, with $R^2 = 0.541$. The positive coefficients for these metrics suggest that an increase in \textit{number of observations} and \textit{temporal occupancy} could effectively mitigate bias.
The regression model suggests that an increase of 100 in the number of observations can reduce the bias by 0.2 stay points. Also, an increase of temporal occupancy by 10 would help reduce the bias by 1.2.
\textbf{Model 2} incorporates three additional data quality metrics but only produced a marginally improved $R^2$. Notably, \textit{percentage of high accuracy observations} reduces bias. In \textbf{Model 3}, \textit{number of observations} and \textit{temporal occupancy} improve $R^2$ yet at the expense of the model's interpretability.
Overall, the metrics \textit{number of observations} and \textit{temporal occupancy} emerge as significant indicators for characterizing the potential bias in LBS data. Meanwhile, the integration of additional data quality metrics and the utilization of more sophisticated models could further refine bias prediction.

\begin{table*}[ht]
    \caption{Results of linear regression models (Dependent variable: Bias on the number of stays)}
    \label{tab:res}
    \centering
    \small
    \renewcommand{\arraystretch}{1.2}
    \begin{threeparttable}
    \begin{tabular}{l@{\hspace{2.5cm}}c@{\hspace{1.1cm}}c@{\hspace{1.1cm}}c}
        \toprule
         \textbf{Independent variable} & \textbf{Model 1} & \textbf{Model 2} & \textbf{Model 3} \\ \midrule
        Constant   & -7.5325 & -13.9969 & -10.7189 \\ 
        Number of observations   & 0.0021 & 0.0016 & 0.0004 \\ 
        Temporal occupancy & 0.1224 & 0.1341 & -0.0934\\ 
        Maximum record gap (min) & & 0.0014 & -0.0011\\ 
        Percentage of high accuracy observations (\%) & & 0.0542 & 0.0491 \\ 
        Temporal burstiness & & 2.0825 & 4.4046 \\ 
        Number of observations (squared) & & & -4.539e-07 \\ 
        Temporal occupancy (squared) & & & 0.0037 \\
        \midrule
        $R^2$ & 0.541 & 0.551 & 0.571 \\ 
        \bottomrule
    \end{tabular}
    \begin{tablenotes}
    \item *All parameters are statistically significant at the 95\% confidence level, with clustered standard errors employed.
    \end{tablenotes}
    \end{threeparttable}
\end{table*}

\section*{Discussion and conclusion}

 This study delved into the quality and inherent biases of LBS data within the Boston Metropolitan Area. Our comprehensive analysis employed data quality metrics and demographic analysis to demonstrate not only the general challenges of discontinuity and sparsity in LBS data but also the significant disparities in data quality across different demographic groups. Furthermore, we introduced a novel resampling-based framework for assessing LBS data bias when studying individual mobility. This framework proves essential for understanding the effects of various LBS data processing methodologies, shedding light on potential biases in LBS data applications and offering insights crucial for developing more accurate and equitable data processing techniques.

Our findings illuminate a striking disparity in LBS data quality, where higher-income, more educated, and predominantly white CBGs display higher sampling rates but suffer from lower data quality, a contradiction likely rooted in heightened privacy concerns. This observation aligns with existing research that suggests socioeconomic and demographic factors significantly influence technology usage and privacy behaviors \cite{porter2006using, park2013digital}, and it is consistent with prior studies highlighting the discontinuity and sparsity of LBS data \cite{he2022percolation, kilic_missingness_2017, groves_shadow_2011}. However, our study extends this discourse by directly linking these behaviors to the quality of LBS data, thereby contradicting the assumption that higher engagement with technology automatically translates to better data quality \cite{huang2018location}. The contribution of this insight is substantial; it challenges the prevailing methods of data collection and processing in LBS studies by highlighting the necessity for nuanced approaches that consider demographic variances. Moreover, this paradox underscores the critical need for developing tailored imputation methods that can address such disparities, thereby ensuring more equitable decision-making and resource allocation across diverse urban populations. Thus, our work not only adds a new dimension to the understanding of LBS data biases but also calls for a reevaluation of current practices in handling LBS data to mitigate the risk of reinforcing existing inequalities.

Additionally, the resampling-based framework developed in this study is important in understanding the effects of LBS data quality. Differing from the existing imputation methods \cite{liu_bidirectional_2021,ren_mtrajrec_2021, gong_high-performance_2020, ugurel_correcting_2024} for mitigating the bias, we focused on quantitatively evaluating the inferential errors arising from varying qualities of LBS data. The significance of this framework lies in its potential to furnish researchers and policymakers with a deeper, more nuanced understanding of biases within LBS data, facilitating more accurate, unbiased conclusions and promoting equitable decision-making processes. This contribution underscores the importance of precise data quality evaluation in the development and implementation of LBS technologies, highlighting a critical step forward in the pursuit of fairness and accuracy in data-driven decision-making.

Our investigation into LBS data within the Boston Metropolitan Area forges new pathways in understanding and addressing data quality and bias, advocating for innovative approaches and methodologies. This research not only enriches the dialogue on demographic impacts on data quality but also sets a foundation for future explorations aimed at enhancing the fairness and accuracy of LBS data applications. Moving forward, the insights garnered here should inspire continued efforts to refine data processing techniques, ensuring that advancements in location-based services equitably benefit all segments of society.

\bibliography{sample}

\section*{Acknowledgements}

The team acknowledges the support from the National Science Foundation (No. 2114197 and 2114260).

\section*{Author contributions statement}

X.W., C.C. and Q.R.W. conceived the experiments, Y.W and E.U. processed the data and helped with literature review, S.H. provided technical support, X.W. conducted the experiments, X.W and Q.R.W. analysed the results and wrote the manuscript, C.C., S.H. and Q.R.W. provided overall guidance and contributed suggestions. All authors reviewed the manuscript. 

\section*{Additional information}

\textbf{Competing interests}

\noindent The authors declare no competing interests.

% To include, in this order: \textbf{Accession codes} (where applicable); \textbf{Competing interests} (mandatory statement). 

% The corresponding author is responsible for submitting a \href{http://www.nature.com/srep/policies/index.html#competing}{competing interests statement} on behalf of all authors of the paper. This statement must be included in the submitted article file.

%%%%%%%%%%%%%%%%%%%%%%%%%%%%%%%%%%
\section*{Data Availability Statement}
%%%%%%%%%%%%%%%%%%%%%%%%%%%%%%%%%%
The data that support the findings of this study are available from Spectus Inc but restrictions apply to the availability of these data, which were used under license for the current study, and so are not publicly available. Data are however available from the corresponding authors upon reasonable request and with permission of Spectus  Inc.

\end{document}